\documentclass[aps,superscriptaddress,preprintnumbers,nofootinbib]{revtex4-1}
\usepackage{graphicx,amssymb,amsmath,color}
\usepackage{hyperref}
\usepackage{multirow}

\newcommand{\beq}{\begin{equation}}
\newcommand{\eeq}{\end{equation}}
\def\bea{\begin{eqnarray}}
\def\eea{\end{eqnarray}}
\newcommand{\bei}{\begin{itemize}}
\newcommand{\eei}{\end{itemize}}

\newcommand{\Fig}[1]{Fig.~\ref{#1}}
\newcommand{\Eq}[1]{Eq.~(\ref{#1})}
\newcommand{\Sec}[1]{Sec.~\ref{#1}}
\newcommand{\App}[1]{Appendix~\ref{#1}}

\def\={\,=\,}
\def\+{\,+\,}
\def\-{\,-\,}

\begin{document}

\title{Localizing Gravitational Wave Sources with Single-Baseline Atom Interferometers}

\author{Peter W.~Graham}
\email{pwgraham@stanford.edu}
\affiliation{Stanford Institute for Theoretical Physics, Department of Physics, Stanford University, Stanford, CA 94305, USA}

\author{Sunghoon Jung}
\email{sunghoonj@snu.ac.kr}
\affiliation{Center for Theoretical Physics, Department of Physics and Astronomy, Seoul National University, Seoul 08826, Korea}
\affiliation{SLAC National Accelerator Laboratory, Stanford University, Menlo Park, CA 94025, USA}


\begin{abstract}
Localizing sources on the sky is crucial for realizing the full potential of gravitational waves for astronomy, astrophysics, and cosmology.
We show that the mid-frequency band, roughly 0.03 to 10 Hz, has significant potential  for angular localization.
The angular location is measured through the changing Doppler shift as the detector orbits the Sun.  This band maximizes the effect since these are the highest frequencies in which sources  live several months.
Atom interferometer detectors can observe in the mid-frequency band, and even with just a single baseline can exploit this effect for sensitive angular localization.
The single baseline orbits around the Earth and the Sun, causing it to reorient and change position significantly during the lifetime of the source, and making it similar to having multiple baselines/detectors.
For example, atomic detectors could predict the location of upcoming black hole or neutron star merger events with sufficient accuracy to allow optical and other electromagnetic telescopes to observe these events simultaneously.
Thus,  mid-band atomic detectors are complementary to other gravitational wave detectors and will help complete the observation of a broad range of the gravitational spectrum.
\end{abstract}


\maketitle

\tableofcontents

\section{Introduction}

LIGO's historic observation of gravitational waves has opened a new spectrum in which to view the universe \cite{Abbott:2016blz,Abbott:2016nmj,Abbott:2017vtc}.  We are already learning much from these first observations (e.g.~\cite{TheLIGOScientific:2016src,TheLIGOScientific:2016wfe}) which raise many interesting questions (e.g.~\cite{Bird:2016dcv,Sasaki:2016jop}).  The next step is to fully exploit the gravitational spectrum by gaining precision information on many gravitational wave sources such as black holes, neutron stars, and white dwarfs.  As LIGO is improved and future detectors such as VIRGO~\cite{ref:virgo} and KAGRA~\cite{ref:kagra} turn on, we will gain ever greater sensitivity to gravitational waves in the high frequency band above about 10 Hz.  But it is also crucial to open up as much of the gravitational spectrum as possible.  Observing different bands in the electromagnetic spectrum led to many important new discoveries, and different bands in the gravitational spectrum seem just as promising.

A major advantage to observing in different frequency bands is their ability to provide a wealth of complementary information (see e.g.~\cite{Sesana:2016ljz}).  For example, an important step to improve gravitational wave astronomy is to improve the ability to localize gravitational wave (GW) sources on the sky.  In order for gravitational wave telescopes to realize their full impact on astronomy, astrophysics, and cosmology they will need to have good angular localization.  This greatly increases the use of GW observations, for example improved angular localization will allow optical and other electromagnetic telescopes to accurately observe the same source, realizing the major goal of multi-messenger astronomy.  As one example, this is necessary to make cosmological measurements with the standard siren program~\cite{Nissanke:2009kt,Cutler:2009qv}.

LIGO will improve angular resolution in the high frequency band in the future, though it is a challenging measurement to make since the sources do not live a long time in this band \cite{TheLIGOScientific:2016wfe, Aasi:2013wya}.  At lower frequencies, sources generally live much longer.  Over a long signal duration, a detector can reorient and the modulation of the observed signal can then give the direction of GW source.  The proposed LISA detector operates around mHz, but even with longer source lifetimes, angular localization is still a challenge for most sources~\cite{Cutler:1997ta, Klein:2015hvg}.  Other more complex, farther-future space missions such as BBO and DECIGO \cite{Takahashi:2003wm,Crowder:2005nr} are expected to achieve better precision~\cite{Cutler:2009qv} with many more satellites.

Atomic interferometry has great potential as a gravitational wave detector \cite{Dimopoulos:2007cj, Dimopoulos:2008sv, Hogan:2010fz, Graham:2012sy, Hogan:2015xla, Graham:2016plp, Geiger:2015tma, Canuel:2017rrp, Chaibi:2016dze, Yu:2010ss, Kolkowitz:2016wyg, Norcia:2017vwu}.  Detectors such as MAGIS \cite{MAGIS} and MIGA \cite{Geiger:2015tma, Canuel:2017rrp} are being planned now.
These detectors are based on  techniques similar to atomic clocks which have  achieved impressive precision \cite{OpticalClock}.  Unlike laser interferometry, atom interferometry (AI) allows sensitive single-baseline gravitational wave detection \cite{Graham:2012sy, Hogan:2015xla}.  Further, it allows observation in the mid-frequency band $\sim$ 0.03 - 10 Hz, between LIGO and LISA \cite{Graham:2016plp}.

\section{Executive Summary}

We will show that the mid-frequencies are ideal for angular localization.  In this band the sources live a long time (usually months at least).  In this observation time, the detector reorients and changes position significantly.  This modulates the observed GW signal in a direction-dependent manner which allows the direction of the GW to be measured with high precision.  Just from observing the gravitational wave with detectors in different orientations we would expect to be able to measure the angular location with an accuracy of roughly $\rho^{-1}$ (in square root of solid angle), where $\rho$ is the signal-to-noise ratio (SNR).  However, with multiple detectors separated by a distance $R$, this angular resolution is enhanced by the ratio $\sim \frac{R}{\lambda}$ where $\lambda$ is the wavelength of the GW.  This enhancement arises because the dominant contributor to the localization accuracy is by measuring the phase lag of the GW between the two detectors.  And in fact, multiple detectors are not necessary, a single detector that orbits over a distance of $R$ during the time of the measurement will have the same enhancement\footnote{For the enhancement, it is necessary that the detector not move in a straight line with constant velocity since this would appear as a constant Doppler shift.}.  Thus in the mid-band where sources live several months, $R$ can be $\sim$ AU and so this enhancement can be several orders of magnitude, allowing good angular localization.

Although a single-baseline detector can have lower cost and risk, it is often questioned whether such a simple configuration can have  precision capabilities and whether a single baseline can provide a measurement of sky position and polarization.  In fact, we will show that mid-band atomic detectors with only a single baseline can provide high precision measurements of the angular location of GW sources as well as other parameters such as polarization or luminosity distance.  So long as the detector baseline reorients rapidly (faster than the source lifetime), it is in many cases equivalent to having multiple baselines in different orientations so the angular location can often be measured  as well with a single baseline as with many.  A terrestrial detector clearly reorients rapidly.  Because a satellite detector can be in an Earth orbit, it can also reorient rapidly with a period of several hours.  Further, within the source lifetime the detector can complete a significant fraction of its orbit around the Sun.  This is equivalent to having multiple detectors spaced by $\sim$ AU and allows the enhanced angular localization discussed above.

Since sources last a long time in the mid-frequency band, this also allows prediction of a merger event a significant time in the future.  Thus, events in this band can be observed and then passed off to LIGO and to electromagnetic telescopes with a prediction of both when and where the merger will occur.  Many sources are localized by the atomic detector with sub-degree resolution, well within the field of view of many optical telescopes.  Thus atomic detectors can provide useful information which is highly complementary to that from LIGO in the higher frequency band, even when observing the same source.

We give details of our calculation of angular localization and other parameters in \Sec{sec:calc}.  Then in \Sec{sec:localization}, we show our results and discuss the underlying physics. We generalize the discussion and  conclude in \Sec{sec:concl}.  Our main results are shown in Tables \ref{tab:benchmark} and \ref{tab:terrestrial} and Figures \ref{fig:accumulation} and \ref{fig:decomposition}.

\section{Calculation of Angular Resolution and Other Parameter Uncertainties} \label{sec:calc}
In this section we give the details of our calculation.  Section \ref{sec: calc detector configs} gives the details of the specific atomic detectors we take as examples.  The actual calculation is discussed in Section \ref{sec calc details}.  The example benchmark signals we consider are discussed in Section \ref{sec calc signals}.

\subsection{Detector Configurations}
\label{sec: calc detector configs}

Our main point is to demonstrate that atom interferometer detectors can provide angular localization of GW sources, even with only a single-baseline.  And further, the mid-frequency band is an ideal band for achieving high angular resolution for many sources.  To illustrate this point we consider two example detectors, a satellite-based detector and a terrestrial detector.  In this paper we are not attempting to actually design detectors, so we simply choose example parameters for these two detectors that are within the possible range.

The satellite-based detector that we consider in this paper consists of two satellites orbiting the Earth at a radius of $2 \times 10^4$ km (orbital period $T_{AI}=7.8$ hours) separated by $130^\circ$.  Each satellite has an AI (free-falling, cold atom clouds) driven by a common laser baseline between the two satellites. The two satellites form a long (single) baseline of $3.62 \times 10^4$ km which is the satellite-based detector.  Since the GW is a plane wave of wavelength much longer than the laser baseline, we simplify the calculation by assuming the GW is measured at the midpoint between the two satellites, orbiting at a radius $R_{AI}=8440$ km. This orbital plane around the Earth is inclined by $\theta_{\rm inc}$ from the ecliptic.

We consider the possibility of having either one or two terrestrial detectors.
Each terrestrial detector operates two AIs vertically separated by $\sim$km, interrogated by a common laser, and forming a vertical $\sim$km-long baseline for measurements of the GW.  We take each detector to be on the equator (just for the sake of computation) and the two detectors are taken separated by $\pi/2$ in longitude; this separation can maximize angular resolution as discussed in \App{app:discrete}.

The satellite detector reorients and repositions quickly by orbiting around the Earth every $T_{AI}=7.8$ hours (orbital radius $R_{\rm AI}=8440$ km) as well as slowly by the Earth's annual orbit around the Sun. Likewise, the terrestrial detector rotates once every day ($T_{AI}=1$ day) and orbits around the Sun every year.  In both cases, the orbit around the Earth reorients more quickly, whereas the orbit around the Sun moves in a larger radius. Consequently, they contain different angular information and become important in different regimes, as will be discussed. The combined orbit around the Sun and the Earth (with a relative angle $\theta_{\rm inc}$) allows detectors to efficiently measure a wide range of directions, helping pinpoint source location and other source parameters (e.g.~polarization) as will be discussed.

\begin{figure}[t] \centering
\includegraphics[width=0.85\textwidth]{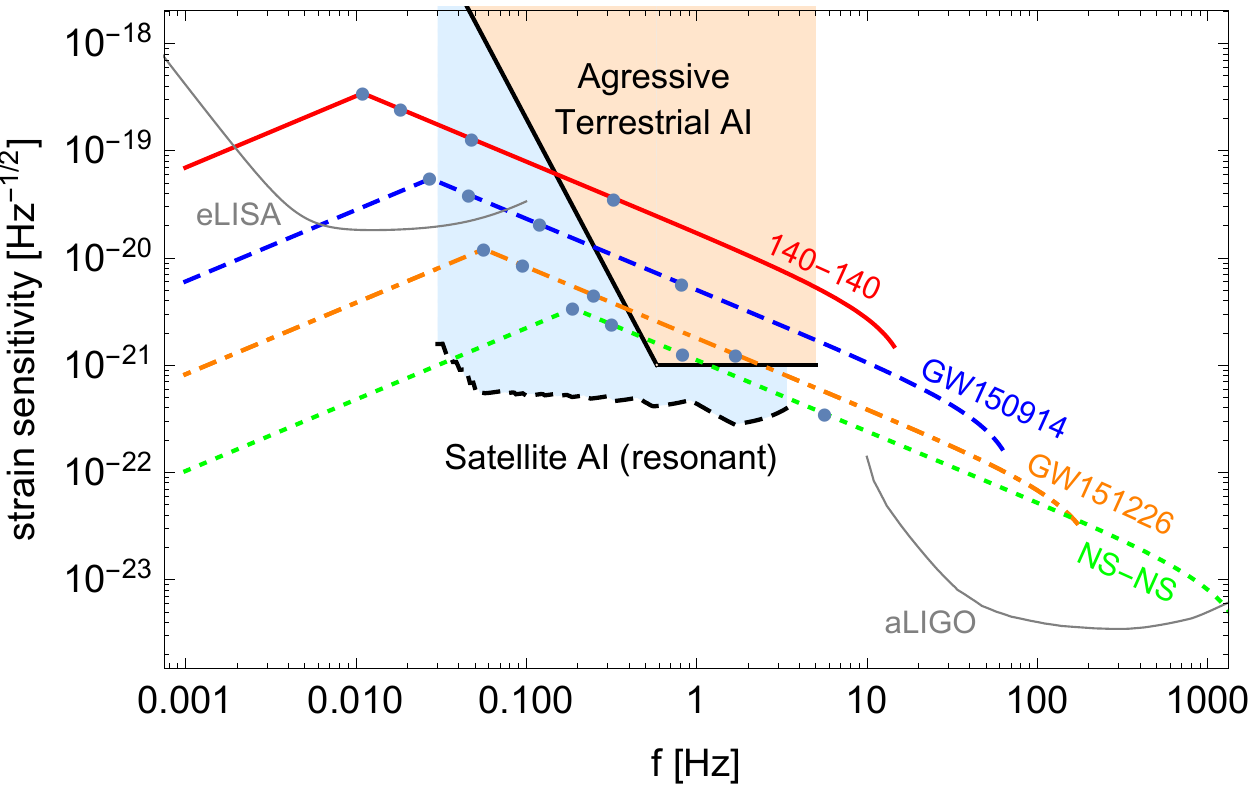}
\caption{Strain sensitivities of the satellite-based resonant AI detector~\cite{Graham:2016plp} and a single aggressive terrestrial AI detector as well as GW strains of benchmarks in Table~\ref{tab:benchmark}: 140-140 (red solid), GW150914 (blue dashed), GW151226 (orange dot-dashed), NS-NS (green dotted). The four dots on each lifetime curve indicate remaining lifetimes of 1 hour, 1 week, 3 months and 1 year until merger. Lifetime curves are bent at 1 year as we consider up to one-year of measurement. The satellite-AI noise is the envelope of the resonant AI detector taken from~\cite{Graham:2016plp}. The sharp increase of the terrestrial noise at low frequencies is due to the GGN; the particular curve is taken from the Homestake mine result~\cite{Harms:2015zma} reduced by a factor 10 as discussed in the text. Also shown are eLISA~\cite{Klein:2015hvg} and aLIGO future design sensitivities~\cite{TheLIGOScientific:2016agk}.}
\label{fig:strain}
\end{figure}

Detector noises that we use in this paper are shown in \Fig{fig:strain}.  Both curves are optimistic projections for future detectors. The satellite detector noise shown is an envelope of a resonant-mode detector discussed in~\cite{Graham:2016plp}, having a $Q$-factor enhanced sensitivity at a certain frequency $f_0$ with a narrow bandwidth $\sim f_0/Q$.  Naturally with this resonant detector the $Q$ factor starts low, $\mathcal{O}(1)$, at the lower frequencies and rises a higher frequencies. Changing the resonance frequency and width can be done in real time by just changing the laser pulse sequence. Thus, a resonant-mode can either be used to track a quickly-evolving, higher-frequency GW source or to simultaneously measure many slowly-evolving, lower-frequency GWs within its narrowband. As an illustration of the angular resolution of AI, we assume that we track each GW source for its last 1 year in this paper; and we use the noise envelope in \Fig{fig:strain} for calculation.  For example, one simple measurement strategy might be to observe in a `detection mode' at a lower frequency (e.g.~around 0.1 Hz) where the resonance is naturally quite broad until a source is detected.  Then that source could be tracked as it evolves upwards in frequency with ever narrowing bandwidth (which should be possible since the SNR is already built up at low frequencies where the frequency of the source did not need to be known well).   We assume a one year measurement of each source just to have a simple calculation with a result that is understandable, but in reality the best way to observe a particular source may be to watch it for only a fraction of each month over a couple years for example, and then use the rest of the time to watch other sources.  The resulting SNR should be about as high as with our calculation, and the angular resolution should be at least as good if not better than what we have calculated.  For good angular resolution, we mainly need several measurements at well-spaced places along the orbit around the Sun, the measurements certainly do not need to be continuous for a year.  See below for more discussion.  In practice, a sophisticated detection strategy for a resonant-mode (switching between tracking a small number of GWs with higher accuracies and observing many GWs) can be developed in future work.

The  terrestrial detector noise shown in \Fig{fig:strain} is based on a broadband-mode. It is an optimistic noise curve that requires advances in the atomic technology that appear possible but have not yet been realized.  This can be taken as a goal, and we will present results for this as it is a benchmark that would be useful to reach.  The sharp increase of the noise at low frequencies is due to  gravity-gradient noise (GGN). The GGN is mainly due to seismic activities and air density/pressure variations, causing  variation of the gravitational potential at the location of the AIs. It is likely to be the dominant background in frequencies below a few Hz~\cite{Hughes:1998pe}. Although it is very challenging to suppress  significantly~\cite{Harms:2013raa}, it can potentially be reduced by various techniques that are being developed, e.g.~from an array of AIs to utilize the fact that wavelengths of GW are longer than the characteristic coherence length of GGN~\cite{Chaibi:2016dze}.  It is also strongly dependent on the location of the detector, so can be reduced by a suitable choice of site and potentially also by other techniques being considered for LIGO  \cite{Harms:2015zma}. We optimistically assume that the  GGN can be improved by a factor 10 from the measurement of Homestake mine noise \cite{Harms:2015zma}; the resulting smaller GGN is shown in \Fig{fig:strain} and will be used in our numerical study.

\subsection{Calculation of Source Parameter Uncertainties}
\label{sec calc details}

A single-baseline detector measures the gravitational stretching and contraction along its baseline direction.  We define the detector response tensor $D_{ij}$ from the baseline direction unit vector $a_i(t)$ as
\beq
D_{ij}(t) \= a_i(t) \, a_j(t).
\eeq
The GW strain tensor is decomposed in terms of polarization tensors $e^{+,\times}_{ij}$ (in \Eq{eq:polarizationtensor}) as
\beq
h_{ij}(t) \= h_+(t) e^+_{ij} \+ h_\times(t) e^\times_{ij},
\label{eq:hij_decomp} \eeq
the observed waveform is given by
\bea
h(t) &\equiv& D_{ij} h_{ij} \= h_+(t) D_{ij}(t) e^+_{ij} + h_\times(t) D_{ij}(t) e^\times_{ij}  \nonumber\\
& \equiv & h_+(t) F_+(t) + h_\times(t) F_\times(t).
\label{eq:waveform-t}\eea
The angular information of the GW source (location, polarization and binary orbit inclination) is encoded in time-dependent antenna functions $F_{+,\times}(t)$ and phases ${\rm arg}(h_{+,\times}(t))$. As a detector reorients and/or moves, the observed waveform and phase are modulated and Doppler-shifted, yielding important angular information. Meanwhile, amplitudes $|h_{+,\times}(t)|$ depend on the chirp mass, luminosity distance and binary orbit inclination. We collect and summarize the GW waveforms $h(t)$ in \App{app:waveform}.

By assuming that the satellite detector orbits around the Earth along $\theta_a = \pi/2$ (where $\theta$ is the Earth's polar coordinate) without loss of generality, we parameterize the detector location on the orbit by a unit vector (in Cartesian coordinates)
\beq
{\bf r}_{\rm AI,0}(t) \= \left( \cos \phi_a(t) ,\, \sin \phi_a(t), \, 0 \right),
\eeq
where $\phi_a(t) = 2\pi t/ T_{AI} + \phi_0$ is the azimuthal orbit angle around the Earth. For the aforementioned satellite-mission orbit, the baseline direction is given by a unit vector
\beq
{\bf a}_0(t) \= \left( -\sin \phi_a(t) ,\, \cos \phi_a(t), \, 0 \right) \qquad \qquad \textrm{(satellite mission).}
\eeq
The vertical baseline of a terrestrial detector is
\beq
{\bf a}_0(t) \= (\cos \phi_a(t) ,\, \sin \phi_a(t), \, 0) \qquad \qquad \textrm{(terrestrial mission).}
\eeq
These vectors defined in the Earth polar coordinate are transformed to the Sun's polar coordinate as (again by assuming the ecliptic at $\theta_{Ea} =\pi/2$)
\beq
{\bf r}_{\rm AI}(t) \= \left( \begin{matrix} \cos \phi_{Ea}(t) & -\sin \phi_{Ea}(t) & 0 \\ \sin \phi_{Ea}(t) & \cos \phi_{Ea}(t) & 0 \\ 0 & 0& 1 \end{matrix} \right)  \cdot \left( \begin{matrix} \cos \theta_{inc} & 0 & -\sin \theta_{inc} \\ 0 & 1 & 0 \\ \sin \theta_{inc} & 0 & \cos \theta_{inc} \end{matrix} \right) \cdot {\bf r}_{\rm AI,0}(t),
\eeq
and similarly for ${\bf a}(t)$. The azimuthal angle of the Earth's orbit around the Sun is $\phi_{Ea}(t) = 2\pi t/ (1 {\rm yr}) + \phi_0^\prime$, and the location of the Earth in the Sun frame is ${\bf r}_{Ea}(t) = (\cos \phi_{Ea}(t), \sin \phi_{Ea}(t),0)$. The inclination $\theta_{inc}$ (of a detector orbit around the Earth with respect to the ecliptic) is chosen to be $\pi/4$ and $23.4^\circ$ for the satellite and terrestrial missions, respectively. (Of course, the satellite-orbit inclination is not a fixed number and its optimal value can be studied.)

\medskip
We estimate parameter measurement accuracies by calculating a Fisher matrix $\Gamma$. 10 free parameters that we consider are binary masses, source direction ${\bf n}={\bf n}(\theta, \phi)$, polarization $\psi$, binary orbit inclination $c_\iota \equiv \cos \iota$, luminosity distance $D_L$, coalescence time $t_c$ and phase $\phi_c$, and spin-orbit coupling $\beta$ (we include $\beta$ in the Fisher estimation although we set $\beta=0$; neutron star spins are small, and spin-orbit dipole interactions are suppressed by large binary separations during the inspiral phase far from merger). The first three angle parameters are defined in the Sun frame. The covariance matrix $\Gamma^{-1}$ is a theoretical estimation of experimental uncertainties. In particular, the angular resolution of the source location is defined as the solid-angle uncertainty~\cite{Cutler:1997ta}
\beq
\Delta \Omega_s \, \equiv \, 2\pi \sin \theta \sqrt{ \Gamma^{-1}_{\theta\theta} \Gamma^{-1}_{\phi\phi} - (\Gamma^{-1}_{\theta \phi})^2}.
\label{eq:angres} \eeq
We use $\sqrt{\Delta \Omega_s}$ (in degree) and the square-root of diagonal elements of $\Gamma^{-1}$ as measurement accuracies in this paper.
More details on Fisher matrix analysis can be found in, e.g. Refs.~\cite{Cutler:1994ys,Cutler:1997ta}, and its utilities and limitations are discussed in, e.g. Refs.~\cite{Vallisneri:2007ev}. Most of our benchmark results have SNR $\rho \gtrsim 5$ so that Fisher matrix can be a good approximation. SNR $\rho$ and Fisher elements $\Gamma_{ij}$ are integrated over the measurement time as
\bea
\rho^2 &=&  4 \int \, \frac{ \tilde{h}^*(f) \tilde{h}(f)}{ S_n(f)} \, df,   \\
\Gamma_{ij} &=&  4 \, {\rm Re} \int \frac{ ( \partial_i \tilde{h}^*) \partial_j \tilde{h} }{ S_n(f) } df,  
\label{eq:gamij}\eea
where $\tilde{h}(f)$ is the Fourier-transform of $h(t)$ as in \Eq{eq:waveform-f}.  If the same data multiples, $\Delta \Omega_s \cdot \rho^2$ is constant.

\subsection{Benchmark Signals}
\label{sec calc signals}

\begin{table}[t] \centering
\begin{tabular}{ c || c  c c | c c c c }
\hline \hline
& & & & \multicolumn{4}{c }{Satellite Detector}  \\
Benchmark  & masses & distance $D_L$ & lifetime  & $\sqrt{\Omega_s}$ [deg] & SNR & $\frac{\Delta D_L }{D_L}$ & $\Delta \psi$ [rad]   \\
 \hline \hline
GW150914 &  36-29 Ms & 410 Mpc & 9.6 months  & 0.16 & 67 & 0.21 & 0.85   \\
GW151226 & 14.2-7.5 Ms & 440 Mpc & 5.5 years   & 0.20 & 16 & 0.88 & 3.4   \\
NS-NS & 1.5-1.5 Ms & 140 Mpc &  140 years   & 0.19 & 5.2 & 2.8 & 11        \\
140-140 &140-140 Ms & 410 Mpc & 25 days   & 0.75 & 190 & 0.80 & 1.2   \\
\hline \hline
\end{tabular}
\caption{Benchmark sources and results for the satellite mission. The most important source parameters are the masses and luminosity distance $D_L$, which determine the overall signal strength and the lifetime spent in the AI frequency band $f = 0.03-5$ Hz. Other source parameters (location, polarization, binary orbit inclination) are somewhat randomly chosen as in \App{app:waveform} (but not leading to particularly good or bad results), and the same set of parameters are used for all benchmarks.  The last four columns show our results for SNR $\rho$ and the uncertainties for angular resolution $\sqrt{\Delta \Omega_s}$ in degrees,  distance $D_L$, and polarization $\psi$ (in radians) from the resonant satellite detector discussed in the text. Results are integrated for the last one year up to 1 hour before merger, or up to the ISCO whichever is earlier.  All uncertainties can be scaled almost linearly with 1/distance. Unphysically large uncertainties (no priors are applied) would mean that the parameters will not be well constrained; but uncertainties linearly-scaled with 1/distance become meaningful and physical for close enough sources.
}
\label{tab:benchmark} \end{table}

\begin{table}[t] \centering
\begin{tabular}{ c || c  c c | c c c c | c c c c}
\hline \hline
& & & & \multicolumn{4}{c |}{One Terrestrial Detector} & \multicolumn{4}{c}{Two Terrestrial Detectors} \\
Benchmark  & masses & distance $D_L$ & lifetime  & $\sqrt{\Omega_s}$ [deg] & SNR & $\frac{\Delta D_L }{D_L}$ & $\Delta \psi$ [rad] &  $\sqrt{\Omega_s}$ [deg] & SNR & $\frac{\Delta D_L }{D_L}$ & $\Delta \psi$ [rad]  \\
 \hline \hline
GW150914 &  36-29 Ms & 410 Mpc & 9.6 months  & 140 & 4.8 & 39 & 140   &  77 & 6.1 & 21 & 70 \\
GW151226 & 14.2-7.5 Ms & 440 Mpc & 5.5 years   & 150 & 1.7 & 45 & 180     &  110 & 2.4 & 31 & 120 \\
NS-NS & 1.5-1.5 Ms & 140 Mpc &  140 years   & 2.6 & 1.1 & 22 & 74       &  1.8 & 1.6 & 15 & 52 \\
140-140 &140-140 Ms & 410 Mpc & 25 days   & 370 & 14 & 94 & 330   &  70 & 14 & 20 & 71 \\
\hline \hline
\end{tabular}
\caption{Benchmark sources are the same as in Table \ref{tab:benchmark}, results shown are for one and two terrestrial detectors. Results are integrated for the last one year up to 10 minutes before merger, or up to the ISCO whichever is earlier. Unphysically large uncertainties (no priors are applied) would mean that the parameters will not be well constrained; but uncertainties linearly-scaled with 1/distance become meaningful and physical for close enough sources. The results with SNR $\rho \ll 5$ may have to be calculated by the Monte-Carlo method as the Fisher matrix may not approximate the minima of the likelihood well, but for closer sources with higher SNR these results could be scaled and would give the correct uncertainties.}
\label{tab:terrestrial} \end{table}

Since two disparate time scales -- $7.8 \sim 24$ hours (earth orbit) and 1 year (solar orbit) -- are present in satellite and terrestrial AI missions, detection strategies and measurement accuracies depend crucially on the GW lifetime in the AI frequency band $f=$ 0.03 - 5 Hz. Although any compact binary mergers lighter than about a few 1000 solar mass will pass AI band, their lifetime in the AI band vary from a few seconds to several years depending on the masses. 

We choose four benchmark GW sources that span a wide range of lifetime in the AI band.  Our point in this paper is not to estimate what sources could definitely be seen with reasonable rates.  So we simply chose some example source parameters.  But it is easy to scale the results for any source to a different distance for example.  Since the strain $h$ of the source scales linearly with distance, then in a wide range of parameters, all the uncertainties we estimate (e.g. for angular resolution, polarization, etc) also simply scale linearly with distance.

The four benchmark sources are shown in Table~\ref{tab:benchmark} and \Fig{fig:strain}. The first two listed are  LIGO's first two discoveries (GW150914~\cite{Abbott:2016blz}, GW151226~\cite{Abbott:2016nmj}), spending at least about a year in the AI band so that both time scales are relevant. Their binary masses and luminosity distances are taken from LIGO measurements~\cite{TheLIGOScientific:2016wfe,Abbott:2016nmj}, but other source parameters such as locations on the sky, orbit inclination, polarization in \Eq{eq:sourceparameters} are chosen somewhat randomly (but we checked that they do not lead to particularly good or bad results; see discussion in \Sec{sec:localization}). However, the masses and distances are the most important factors determining signal strengths and frequencies. In addition to the LIGO sources spending at least about a year, we consider a  neutron star binary (NS-NS).  This is an important source of GWs that is expected to produce electromagnetic signals during merger as well.  We chose it at a distance which LIGO could see as well, and in which it is reasonable to expect at least some events during a mission lifetime.  Lastly, we consider a black hole (BH) binary with two 140 solar mass BH's, ``140-140'', which spends only 25 days in the AI band, representing the case where only the smaller time scale -- $7.8 \sim 24$ hours -- is relevant.  This source would likely not be visible at LIGO since it merges at too low a frequency.  This type of source is not known to exist, but such heavy BH's would likely only be seen when we turn on a GW detector in this mid-frequency band anyway.  So here we have simply chosen as an example distance, the same distance as LIGO's first source, but the results can be easily scaled to any other distance as discussed.
Results will be discussed in \Sec{sec:localization}.

\section{Results for Angular Localization and Other Source Parameters} \label{sec:localization}

In this section we discuss the results of our calculation for the angular localization and uncertainties in other parameters such as polarization and luminosity distance.  We give an approximate analytic understanding in Section \ref{sec analytic results}.  Our specific numerical results are discussed in Section \ref{sec specific results}.  The robustness of our results against varying detector and source parameters is discussed in \ref{sec parameter dependence}.

\subsection{Approximate Analytic Description}
\label{sec analytic results}

The single-baseline AI measurement contains directional information in the form of the modulation of GW signal strength and phase as the AI detector reorients and orbits. When the detector baseline reorients, the relative angle between its baseline and source direction (and polarization) changes so that the observed signal strength (signal-to-noise ratio SNR $\rho$) and polarization-phase (in  \Eq{eq:polphase}) changes. Thus the angular resolution from such information improves roughly as $\sqrt{\Delta \Omega_s} \propto 1/ \rho$. Moreover, if a detector orbits (or, moves non-linearly, in general), the Doppler-shift of the GW phase (in \Eq{eq:dopplerphase}) changes. As the phase accumulation changes linearly with frequency $f$ and orbit radius $R/c$, such angular resolution improves approximately as $\sqrt{\Delta \Omega_s} \propto 1/ (\rho \, 2\pi f R/c)$. By roughly comparing the two contributions, we expect that the Doppler effects would dominate when $fR/c \gtrsim 1$ (or $f\gtrsim 10^{-3}$ Hz for $R=$ 1 AU $\simeq$ 500 sec/c), improving angular resolutions in proportion to the GW frequency. Thus the interplay of the two effects will determine overall performance and properties of atomic measurements, as will be discussed below. We will call the two effects by ``reorientation'' and ``Doppler'' effects; but the latter effect is actually the \emph{change} in the Doppler effect over time.  These give an approximate analytic description of the enhancement in angular resolution.

The above estimations are valid only when full angular information is gained by measuring at a large range of angles. Consider the Doppler effect from a circular orbit. The GW phase will be periodically Doppler-shifted and will modulate along the orbit, giving important information about source direction. But a small portion of the orbit  is close to linear motion so that Doppler effects do not change much. Since the linear Doppler shift is not measurable (redshift is not measurable from GW measurements since we do not know the source frequency~\cite{Nissanke:2009kt}), exactly linear motion with a constant velocity would not give us any angular or polarization information.  So a small portion of an orbit alone does not provide good angular resolution. This means that, in order for the Doppler effect around the Sun to be fully utilized, a GW should spend at least a few months in the AI band (and most benchmarks indeed do). 


\begin{figure*}[t] \centering
\includegraphics[width=0.49\textwidth]{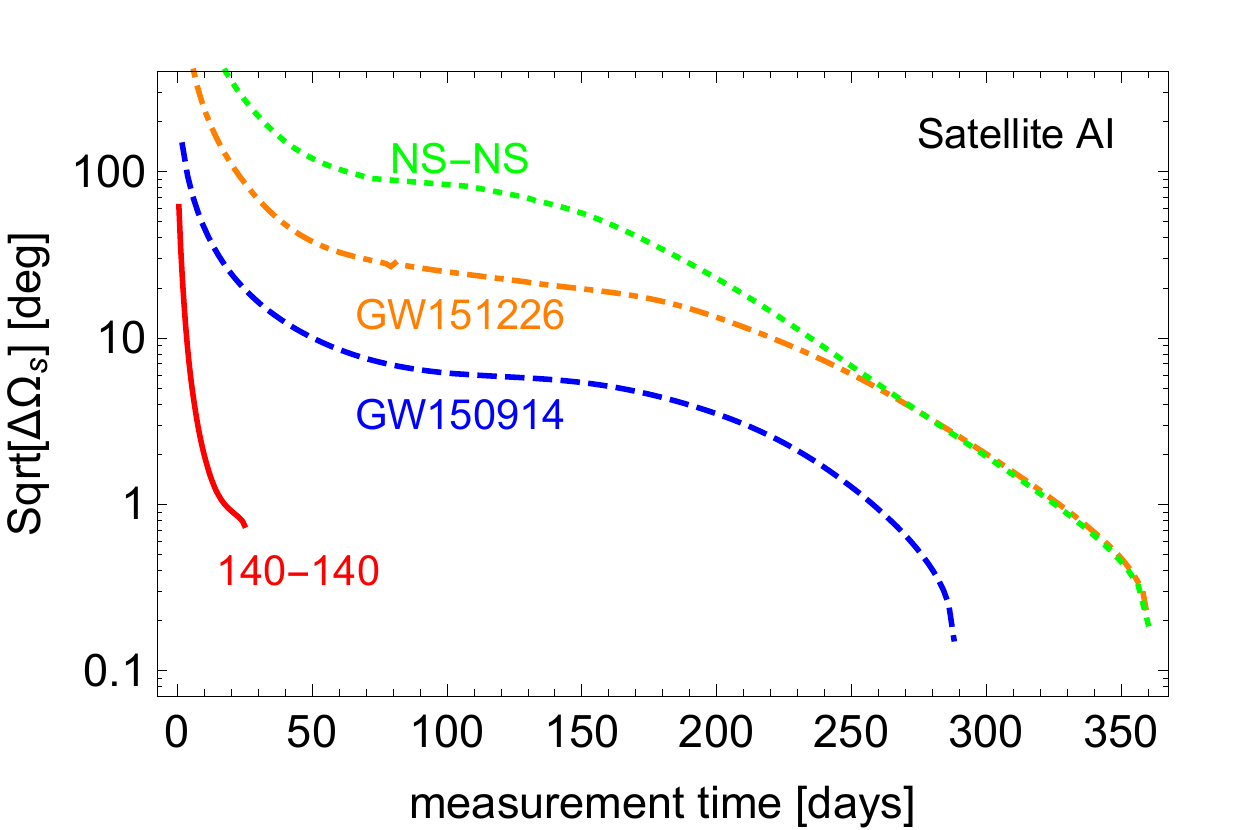}
\includegraphics[width=0.49\textwidth]{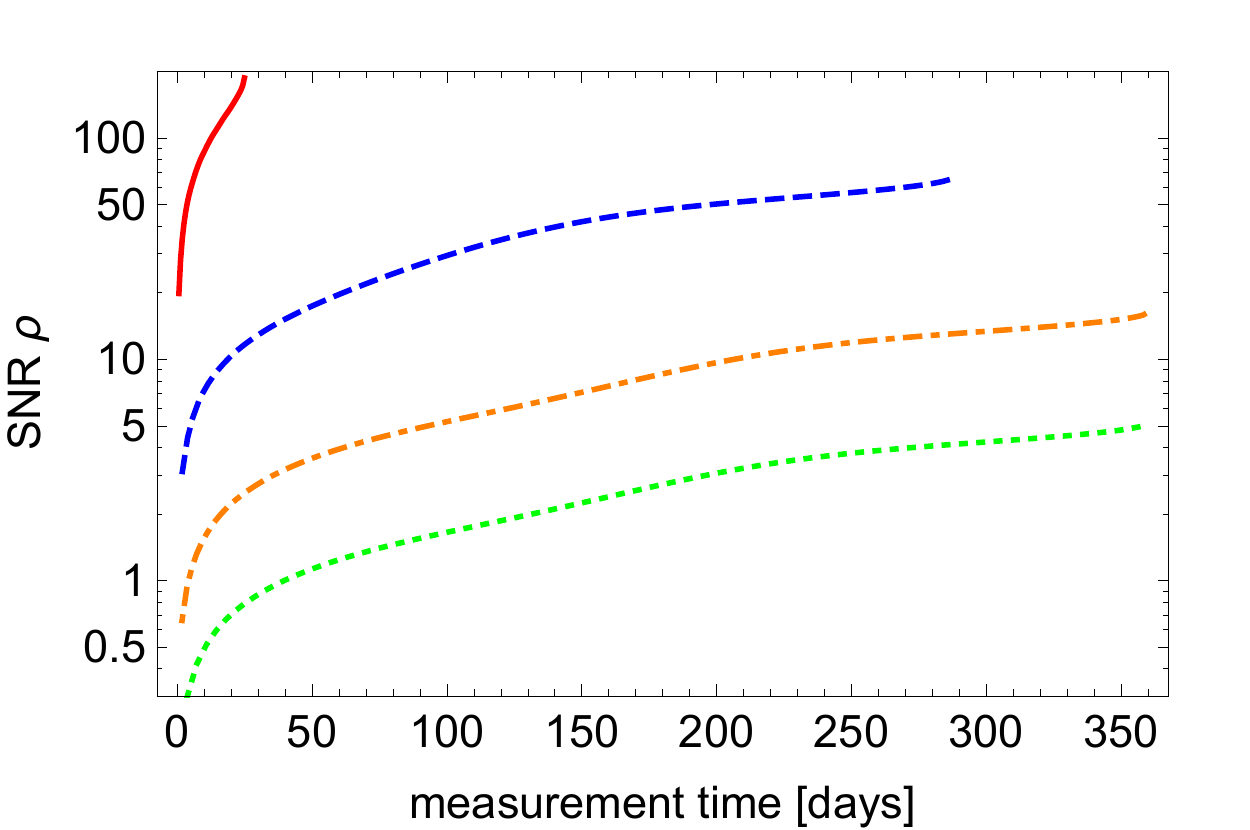} \\
\includegraphics[width=0.49\textwidth]{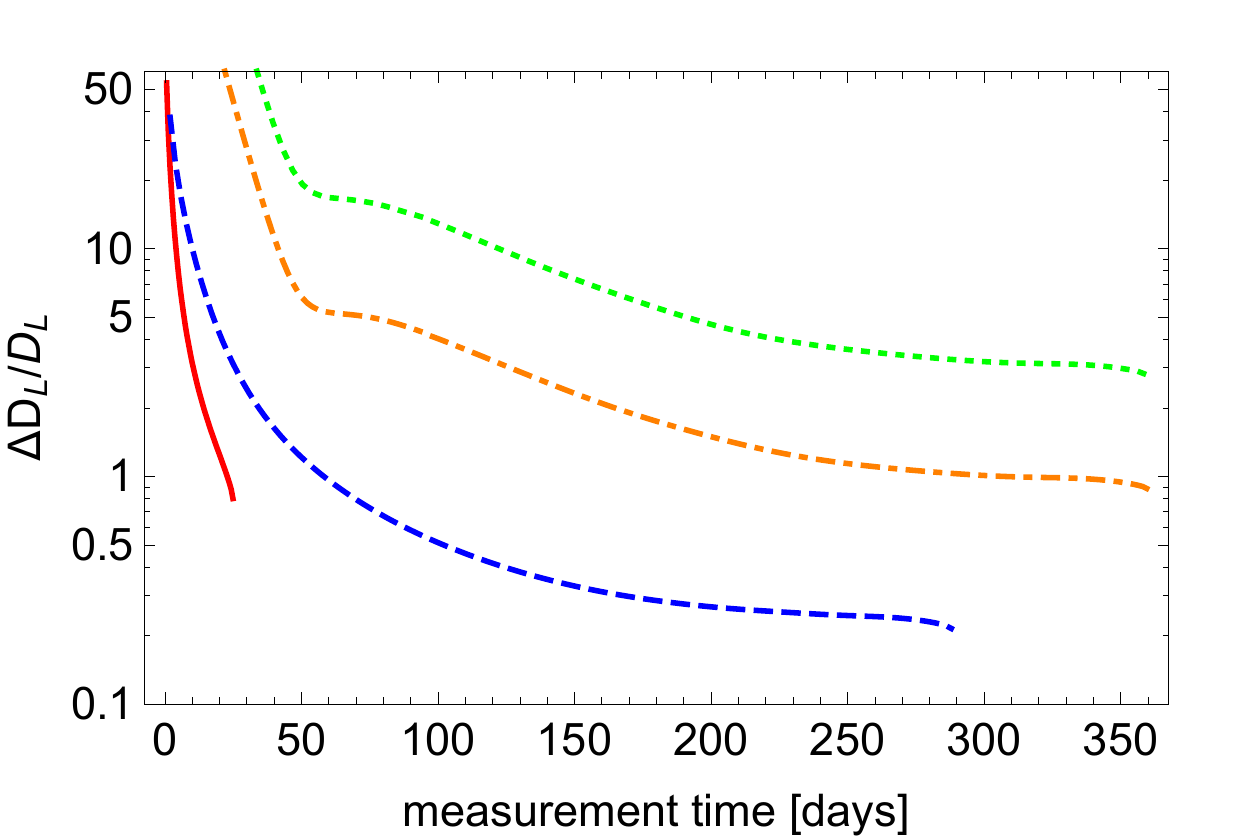}
\includegraphics[width=0.49\textwidth]{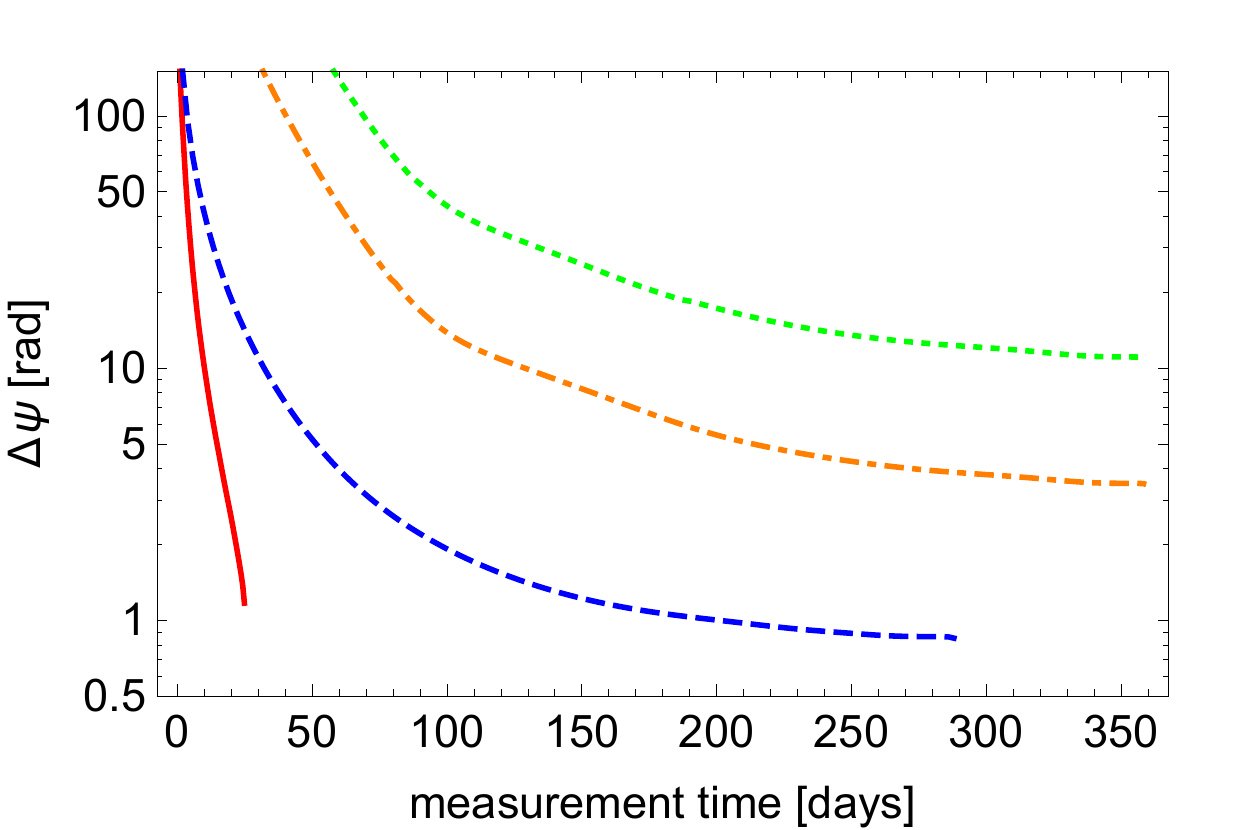} \\
\includegraphics[width=0.49\textwidth]{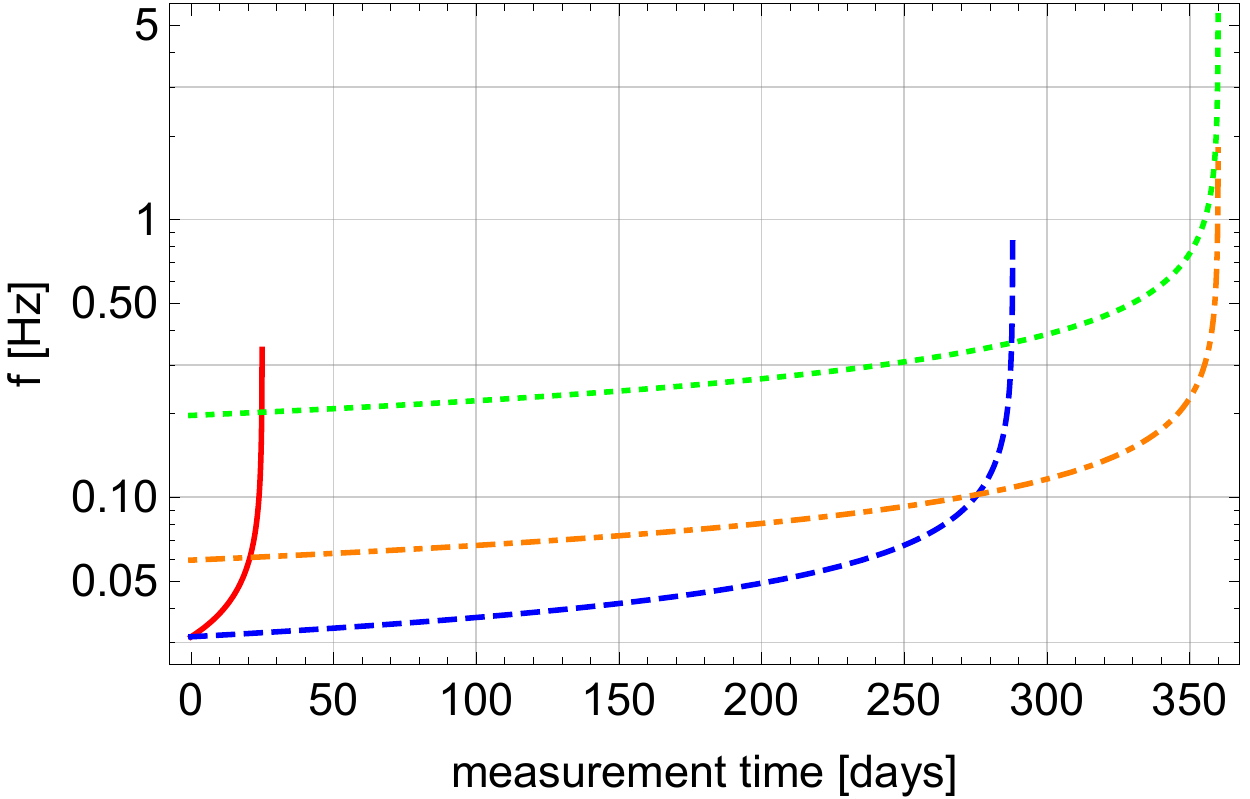}
\caption{Accumulation of measurement accuracies from the satellite resonant-AI detector for the benchmarks in Table~\ref{tab:benchmark}: 140-140 (red solid), GW150914 (blue dashed), GW151226 (orange dotdashed), NS-NS (green dotted). The observables shown are location angular resolution $\sqrt{\Delta \Omega_s}$, SNR $\rho$, errors on luminosity distance $\Delta D_L / D_L$ and on polarization $\Delta \psi$ as functions of measurement time starting from 1 year before merger (or as soon as the GW enters the AI band) up to the last 1 hour or the ISCO, whichever is earlier. No priors are added. Errors and SNR improve almost linearly with 1/distance for the given source parameters.  The bottom figure is the source frequency as a function of time for reference.}
\label{fig:accumulation}
\end{figure*}
\begin{figure*}[t] \centering
\includegraphics[width=0.49\textwidth]{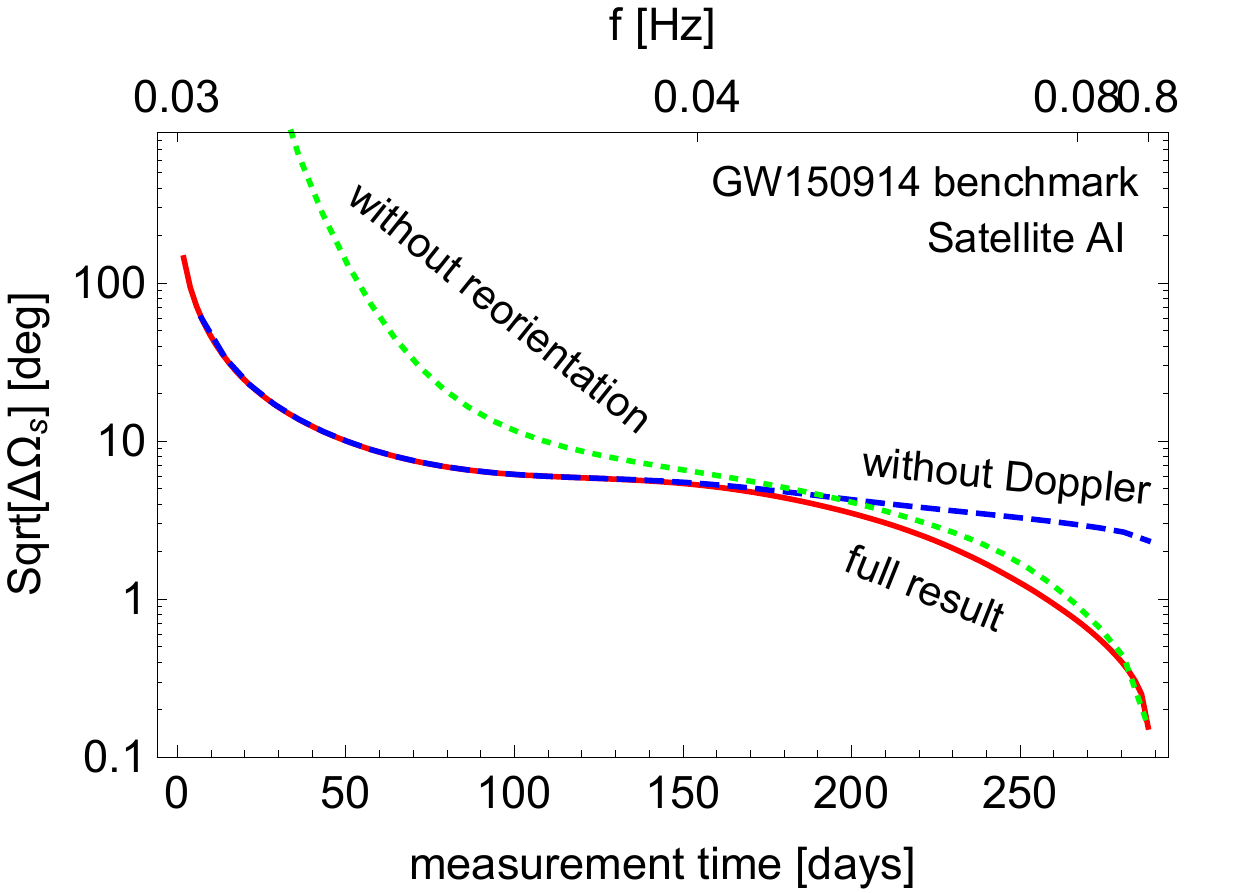}
\includegraphics[width=0.49\textwidth]{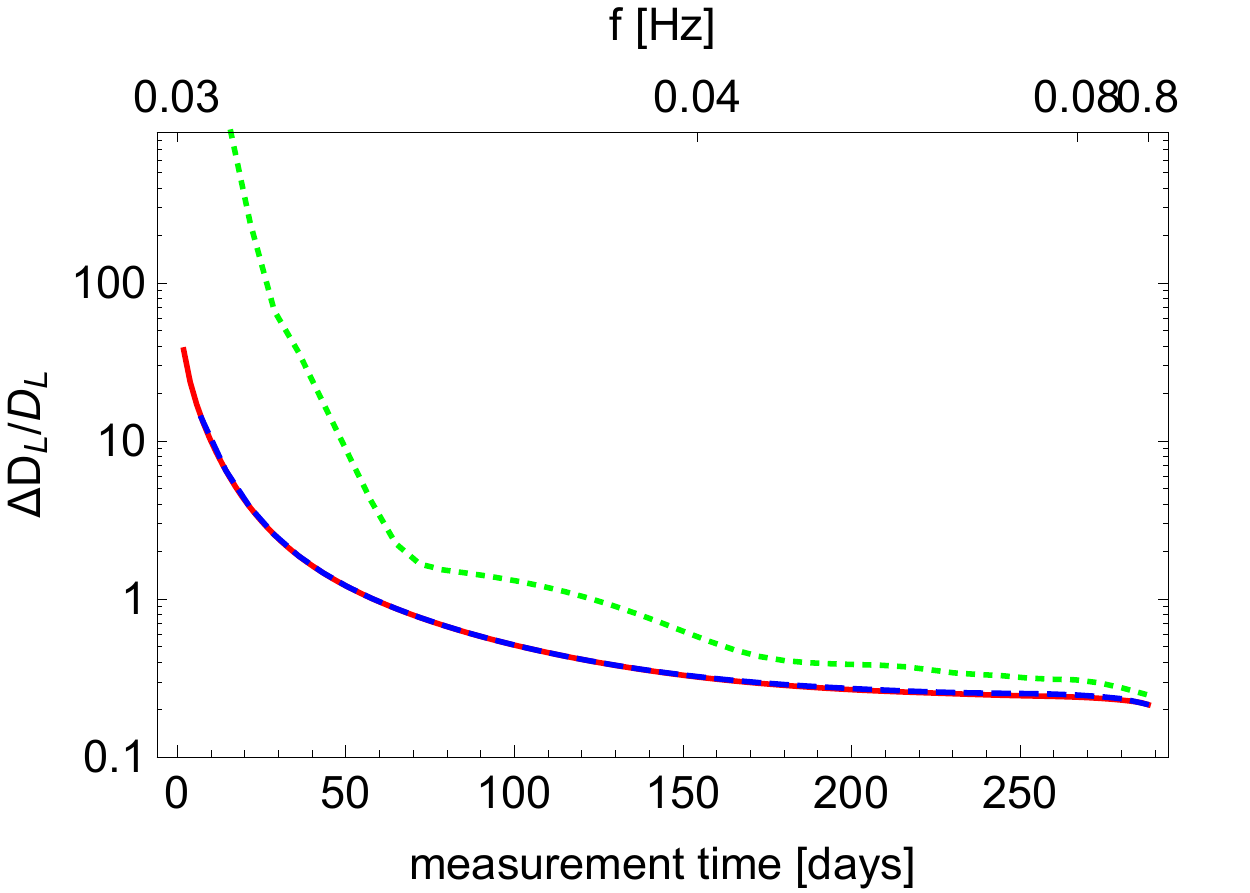}
\caption{Decomposition of angular resolution $\sqrt{\Delta \Omega_s}$ and $\Delta D_L/D_L$ by reorientation and Doppler contributions.   The source is the GW150914 benchmark as seen in the satellite detector. Shown are full results (red-solid), Doppler effects ignored (blue-dashed), or detector baseline not reorienting in the Earth frame (green-dotted). The left panel illustrates that reorientation is important in the first few months, but Doppler effects can further improve angular resolution after a few months. The right panel shows that $D_L$ (and other source parameters) can only be well measured by detector reorientation; in the green-dotted curve, the AI baseline is fixed in the Earth frame leading to poor resolution until the detector slowly reorients by orbiting around the Sun.}
\label{fig:decomposition}
\end{figure*}

We first demonstrate the underlying physics in \Fig{fig:accumulation} and  \Fig{fig:decomposition}. The results are obtained by integrating the last one year of  AI measurement (or as soon as the GW enters the AI band) up to 1 hour (for satellite mission) or 10 minutes (for terrestrial mission) before merger. This time gap is to allow some time to warn LIGO and telescope follow-ups; in later sections, we vary the time gap. Compact binaries with masses $M_1$ and $M_2$ are assumed to merge at the innermost stable circular orbit (ISCO) of the orbital separation $r= 6 M_z = 6(M_1+M_2)(1+z)$ corresponding to $f=(6\sqrt{6} \pi M_z)^{-1}$. Thus, we conservatively include only the inspiral stage but not the merger and ringdown (although they can be important for very short signals merging in the AI band). Based on these and our Fisher estimations, \Fig{fig:accumulation} plots the accumulation of angular resolution$\sqrt{\Delta \Omega_s}$ in \Eq{eq:angres}, SNR $\rho$, the uncertainty of the luminosity distance $\Delta D_L/D_L$ and polarization $\Delta \psi$ as functions of the measurement time. \Fig{fig:decomposition} decomposes the contributions to $\sqrt{\Delta \Omega_s}$ and $\Delta D_L/D_L$ from reorientation and Doppler effects for GW150914 benchmark.

The first panel of \Fig{fig:accumulation}  shows the discussed interplay of Doppler versus reorientation  for the case of satellite AI measurements. Angular resolutions generally improve quickly in the first few months, then reach a plateau, and improve again in the last few months. 
In the first few months, most angular information is gained from the quick reorientation around the Earth ($T_{AI}=7.8$ hours). But after a few months, this effect is saturated and becomes dominated by the Doppler effect around the Sun, improving the angular resolution by another few orders of magnitudes. The growth of angular resolution near merger is particularly prominent as frequency chirps quickly (so does Doppler effect). Meanwhile, the 140-140 in \Fig{fig:accumulation} shows only the initial growth of angular resolution as it spends too short time in the AI band. The decomposition of two main effects is demonstrated in \Fig{fig:decomposition} left panel. The intermediate plateau in the figure is the result of the saturation of reorientation effects before Doppler effects begin to dominate. 

Although Doppler effects predominantly determine the angular resolutions of long enough GWs, the reorientation effect is still essential for the precision measurement of other source parameters. We show measurement accuracies of $D_L$ and $\psi$ in \Fig{fig:accumulation}; although we show only these two, we checked that most discussions here apply to other important parameters such as orbit inclination $c_\iota$ as well.
The \Fig{fig:accumulation} shows that $\Delta D_L/D_L$ and $\Delta \psi$ accuracies grow significantly only in the first few months because of reorientation effects but do not grow very much after that by Doppler effects. Supporting this, \Fig{fig:decomposition} right panel shows that the full result (red-solid) and the result without Doppler effects (blue-dashed) almost coincide; but if reorientation is approximately ignored (green-dotted), the resolutions of such parameters are degraded significantly. This result is understood because a Doppler phase in \Eq{eq:dopplerphase} depends only on the flight-time difference or the sky location ${\bf n}$ projected along the displacement but not on other source parameters. 
This study can provide important feedback for designing a real satellite mission, for example in choosing orbital parameters to maximize the ability to measure all the source parameters.

It is remarkable that a single-baseline detector can utilize all these physics. Although multi-baseline measurements can add more information on various parameters, the essential part of the underlying physics -- the change of Doppler effect -- is induced (regardless of detector details) by a non-linear trajectory of the detector, which is provided by the rapid orbit around the earth and the annual orbit around the Sun.  Our satellite mission is indeed designed to contain a rapid orbit around the Earth ($T_{\rm AI} = 7.8$ hours); and terrestrial detectors automatically reorient every day.

\subsection{Specific Results for Single Baseline Detectors}
\label{sec specific results}

We now turn to a discussion of specific results for the satellite and terrestrial detectors.  Our main results are tabulated in Tables \ref{tab:benchmark} and \ref{tab:terrestrial}. Foremost, the satellite mission's mid-frequency measurements can be extremely useful and unique. Its angular resolution ${\cal O}(0.1)$ deg with SNR $\sim {\cal O}(10)$ is comparable to typical field-of-views (FOV) of ground-based telescopes $\sim 1$ deg. This enables such telescopes to search for electromagnetic follow-up signals. Since the resolution scales almost linearly with 1/distance, similar sources within about ${\cal O}(10)$ Mpc can  even be followed up by the Hubble telescope (FOV $\sim 0.007-0.05$ deg). The electromagnetic identification is the first step of multi-messenger physics and standard-siren program~\cite{Nissanke:2009kt}. It is also worthwhile to note that the sub-degree resolution is achieved with relatively small SNRs (just big enough for discovery). For comparisons, LISA measuring lower frequencies with three detector baselines can achieve 0.5 deg resolutions with SNR $\sim 100-10000$~\cite{Cutler:1997ta}, and LIGO-band measurements with many detectors can rarely find such well-localized events~\cite{Aasi:2013wya}. This means that the mid-frequency band has a nice economical balance of high frequency (for large Doppler shift) and long lifetime (for large change) for good angular resolutions with small SNRs. 

We emphasize that the first two benchmarks are based on actual (measured) properties of  GW sources observed by LIGO. The sub-degree resolution means that those discovered GWs could have been localized and warned in advance from the mid-frequency measurement.

On the other hand, the terrestrial detector's angular resolutions are about two orders of magnitude worse. Other source parameters are also not so well measured\footnote{Their uncertainties are unphysically large as we do not add physical priors to the Fisher matrix. But we still show them because uncertainties scale with 1/distance, and the scaled results can be meaningful for close enough sources.}. It is mainly because the low-frequency regime is swamped by GGN so that only the short measurement of the high-frequency regime becomes useful (such useful durations are from a few hours to a few days; see \Fig{fig:strain}). Consequently, no full Doppler effects can be utilized; we recall that the change of Doppler effects measured over several months is the one that enhances angular information. Although the results can perhaps be used by the Fermi satellite (FOV $\sim 90$ deg), reducing GGN in mid-frequency terrestrial detectors is one of the most important tasks for a terrestrial experiment~\cite{Harms:2013raa}.

We also learn by comparing among benchmarks. A notable result of the NS-NS localization $\sqrt{\Omega_s}$ is that it can be done much better than other benchmark localizations for the given SNR. 
It is because the NS-NS spends longer time in high-frequency band that subsequently enables larger Doppler effects. This contrast is most clearly seen in our terrestrial results as only the high-frequency regime is useful there. 
By the same reason, on the other hand, the 140-140 spends too short time in the high-frequency band so that its localization is not significantly better in proportion to its high SNR, $\rho$. Thus, angular resolutions among different GW sources do not simply scale as $\sqrt{\Omega_s} \propto 1/\rho$, but depend also on the frequency content of the GW, and the motion of the detector during the observation time.

The uncertainties of $D_L$ and $\psi$ (and $c_\iota$ as well) from the satellite-mission results in Table~\ref{tab:benchmark} follow the 1/SNR scaling better among GW150914, GW151226 and NS-NS results (except the 140-140). For example, $\Delta D_L /D_L$ of NS-NS is about 20 times worse than that of GW150914 as its SNR is about 20 times smaller, and so on. 
 Here, we find that the localization and measurements of other source parameters become decoupled. The former is dominantly determined by Doppler effect, whereas the latter is by reorientation. In addition, the reorientation effects for GW150914, GW151226 and NS-NS are saturated. Therefore, measurements of other source parameters improve simply with  data, i.e., 1/SNR. 
 This also explains why terrestrial results of $\Delta D_L/D_L, \Delta \psi$ (and $\Delta c_\iota$) do not scale with 1/SNR among benchmarks. Doppler effects are not fully utilized in these cases so that uncertainties and correlations are general mixtures of (unsaturated) Doppler and reorientation effects, obscuring any simple scaling rule. Quantitatively, a dimensionless quantity measuring the correlation
$\Gamma^{-1}_{ij} / \sqrt{ \Gamma^{-1}_{ii} \Gamma^{-1}_{jj} }$
for $i \in \{\theta, \phi \}, \, j \in \{ D_L, \psi, c_\iota \}$
is ${\cal O}(0.1)$ for the terrestrial results, which is 5-10 times larger  than that of the satellite mission.

From our results, we can see the use of having multiple detectors.  For example, the results for one and for two terrestrial detectors are shown in Table \ref{tab:terrestrial}.  Note that for the very long-lived sources (GW151226 and NS-NS) all results improve by a factor of $\sqrt{2}$ from having two detectors instead of one.  This makes sense because for a source that lives a long time and is not changing much (compared to the reorientation and orbit times of the detector), having one detector that moves is basically equivalent to having multiple detectors.  However we can see that the shorter the observation time, the bigger the effect of having multiple detectors.  For example, for GW150914 the improvement is almost a factor of 2.  For the 140-140 source the improvement is even bigger since its lifetime in our band is quite short.  This trend makes sense since a shorter lifetime in band gives the single detector less time to move.  This effect is amplified for our terrestrial detectors because the large noise at low frequencies means that it is mainly the higher frequencies that are useful for parameter estimation.  The sources spend a short amount of time in this band, for example for the 140-140 source it is mainly the last hour of lifetime that determines our calculated uncertainties.  In this time, a single terrestrial detector has only small changes in its position and orientation, thus minimizing its ability to measure these source parameters.  Although we do not show the results, we checked that this same pattern holds true for having multiple satellite detectors.  In that case, since the sources live a long time, for most sources there is little gain from having two detectors (other than the factor of $\sqrt{2}$).

We have not discussed white dwarf (WD) binary mergers here, although those events, which could presumably give a Type Ia supernova for example, are certainly interesting.  They would be interesting and important to observe in gravitational waves and might be observable in this intermediate frequency band.  However we have left discussion of these to future work because any observation in the intermediate frequency band would be when the binary is near merger and there are significant non-gravitational effects on the binary and the GW waveform (see e.g.~\cite{Paschalidis:2009zz, Paschalidis:2011ez, Shen}).

\subsection{Dependence on Detector and Source Parameters}
\label{sec parameter dependence}

As a very preliminary step towards optimizing measurement protocols, we study the impact of varying the measurement time.  As mentioned above, we take the measurement time for each source to be the last year of its lifetime for simplicity and so our results are easily interpretable.  But in reality the optimal observation strategy is likely to be quite different.
As mentioned above, the optimal observing strategy may be to detect sources at the lower frequencies and then keep coming back to observe them periodically as they rise to higher frequencies.  This lets us observe them at different parts of the detector's orbit around the Sun, maintaining the $\sim f R$ enhancement to the angular resolution, and helping in the final prediction for merger time, but still allows observing time to watch other sources as well. 
For example, our benchmark NS-NS binary source is seen with an SNR of 5.2 in the last year.  However if the preceding one year is used instead (i.e.~the second to last year of the binary's life), the SNR is 3.3.  So not much SNR would be lost by using some combination of observing times over the last few years of that source's life.  
The precise optimal observing strategy including spacing of measurements for each source is beyond the scope of this paper (but see Appendix \ref{app:discrete} for a few more details).  We simply note that in order to attain the angular resolution enhancement we have discussed, it is necessary to observe the source from a few different points spaced by $\mathcal{O}(1)$ around the detector's orbit around the Sun.

As another example we consider the effect of taking the measurement of each source to end one day before the merger (allowing more time for follow-up observations to be ready) instead of one hour or 10 minutes. For satellite-mission results, the angular resolutions are the most affected and mildly worsen by a factor of 1.5-3, whereas the other parameter measurements are affected even less. This pattern of impact can be understood from \Fig{fig:accumulation} by cutting the final one day of the measurement time. The main change of the angular resolution results from the reduction of Doppler effects that are still growing rapidly at the end of the measurement time; in contrast, reorientation effects are well saturated by that time, and so do not lose much information from losing the final day of observation. These mild sensitivities to measurement time can be helpful in designing detection protocols. For terrestrial-mission results, on the other hand, all measurement accuracies degrade by an order of magnitude. The degradation is well captured by the loss of SNR by a factor 6 or so (satellite-mission's SNRs reduce only by 10\%). As the useful measurement times (which are not swamped by GGN) are short, the reduction of measurement time causes a bigger loss of measurement accuracies. Thus, reducing GGN will again improve this situation.

So far, we have chosen and fixed one particular set of source parameters. How would the results change with different source parameters? 
We varied ${\bf n}, \psi$ and $c_\iota$ to obtain the possible range of measurement accuracies. It turns out that the 140-140 result and terrestrial results are most sensitive to the choice of source parameters; accuracies vary generally by a factor 10 up or down. On the contrary, results for the other sources in the satellite-mission  vary generally only by 20-30\% for GW150914, GW151226 and NS-NS. The big sensitivity  stems mainly from the short measurement time. The short measurement time means the detector can essentially span only a 2-dimensional plane.  Whereas in the satellite mission, the long measurements allow the detector to cover a significant part of its orbit around the Sun meaning the single baseline sweeps out a 3-dimensional volume. Thus, for short measurement times, particular source angles with respect to the 2-dimensional plane play a crucial role in the observed signal strength. Though it is possible that different choices of orbit around the earth could improve this situation for the short measurements.  If it were not for rapid reorientation, the parameter dependencies would have been terrible; but these single-baseline AI detectors naturally reorient as discussed.  In general, the assumption we have made that the satellite detector orbits the earth can play a crucial role for several types of sources by allowing rapid reorientation of the detector.

\section{Conclusion} \label{sec:concl}

In order to fully realize the promise of gravitational wave astronomy, we will need the ability to accurately localize detected objects on the sky.  Angular localization is a crucial feature of any telescope, and for example will allow optical and other electromagnetic telescopes to  observe the same source, greatly increasing the information gained.

We have demonstrated that the mid-frequency band, roughly 0.03 to 10 Hz, is in many ways the ideal band for the angular localization of many gravitational wave sources.  Even sources that are not observed with very high SNR can nevertheless be localized to high precision on the sky (see Tables \ref{tab:benchmark} and \ref{tab:terrestrial}).  The angular resolution is enhanced (approximately) by the ratio of the distance over which the detector moves during the measurement of the source (or the distance between multiple detectors) to the wavelength of the GW.  For the mid-frequency band this can be an enhancement of some orders of magnitude.  In this band, most sources live at least a few months, allowing the detector to move over roughly an AU.  Of course, sources live even longer at lower frequencies but this does not help since the earth never moves farther than 2 AU\footnote{The motion of the solar system through our galaxy, or of our galaxy through the universe, does not help with localization or other source parameter estimation.  This motion is in a straight line with constant velocity and so is a constant Doppler shift.  This kind of motion is not equivalent to having simultaneous, multiple detectors spread over that distance and  does not improve angular localization.}.  And at lower frequencies the wavelength becomes longer, reducing the effect.  For example in the mHz band (LISA's band) the GW wavelength is about an AU, significantly reducing this enhancement to the angular resolution.  So the mid-frequency band appears optimal for angular localization since these are the highest frequencies in which sources live several months.

This is a general point that applies to any type of gravitational wave detector, not just the atomic detectors we have discussed.  The power of the measurement comes from the large change of the Doppler shift over a measurement time of several months.  This is induced not by multiple detectors but by the Earth's (and hence the detector's) orbit around the Sun.  Therefore any detector with good sensitivity in the mid-band ($\sim$0.03 to 10 Hz) would benefit from the improved angular localization we have discussed.  Thus by observing in the mid-band, sources can be localized well before merger, and a precise location and time of the merger event can be predicted, allowing optical and other electromagnetic telescopes to observe the merger event simultaneously.

Specifically, single-baseline atomic detectors in this band can provide excellent angular localization, often sub-degree, for many important sources, see Table \ref{tab:benchmark}.  These single baseline detectors can also provide good measurement of the other source parameters such as the GW polarization or luminosity distance.  Because the detector reorients and moves significantly during the source lifetime, a single baseline is all that is needed for this precision measurement of the source parameters.  Having multiple baselines, or multiple detectors, does not significantly change the precision of the measurement of any of the parameters (the angular location, polarization, distance, etc.).  A single baseline detector that reorients and repositions can be as good as multiple detectors\footnote{So long as it does not move in a straight line with constant velocity, as discussed above.}.  Multiple baselines or multiple detectors are certainly not required for angular localization or measurement of other parameters.  This allows atomic detectors to provide useful information that is  complementary to other detectors such as LIGO or LISA.
In fact,  having several of these detectors observe the same source across a wide range of frequencies may  allow measurements significantly better than any one of them could achieve on its own.
It would be interesting to study the gain from observing a source by multiple detectors (see e.g.~\cite{Sesana:2016ljz}), particularly if the same source can be observed by both atomic detectors and LIGO or LISA.

The  satellite atomic detector considered here has excellent angular resolution for a gravitational wave telescope.  In fact a discovery almost guarantees sub-degree angular resolution. And many sources can be localized down to ${\cal O}(0.1)$ degree.  A terrestrial atomic detector could possibly give useful angular information with future improvements in technology, especially in the ability to reduce GGN.  Strategies to subtract such noise using a number of AIs~\cite{Chaibi:2016dze}, careful choice of site \cite{Harms:2015zma}, seismometers around a detector~\cite{Harms:2013raa,Driggers:2012ac}, and measurement of gravity-gradient tensor components~\cite{Harms:2015zaa} are being developed.  The fact that the satellite detector can orbit the Earth (and the terrestrial detector `orbits' Earth as well) is useful because it leads to rapid reorientation of the detector which improves measurement of parameters such as the polarization and helps remove degeneracies with the angular resolution.  We  chose particular orbits as an example, but this calculation should in fact be used to inform the choice of orbit when optimizing the design of a real satellite mission.

The discovery of gravitational waves has opened an entirely new window to our universe.  The next step is to exploit this ability as fully as possible.  This will involve observing new parts of the gravitational spectrum, in particular at lower frequencies, and gaining as much precision information as possible from these observations.  The mid-band, frequencies just below LIGO's,  allows excellent angular localization, as well as allowing early warning of events such as black hole, neutron star, and white dwarf mergers.  This will allow us to exploit the power of multi-band and multi-messenger astronomy, combining simultaneous electromagnetic and gravitational measurements to enable a deeper study of the universe.

\begin{acknowledgments}

We thank Jason M. Hogan for many useful discussions on detector configuration and noise curves and Rana Adhikari and Leo Singer for various comments and checking our antenna functions for single-baseline detectors.  We also thank Douglas Finkbeiner, Mark Kasevich, Peter Michelson, Surjeet Rajendran, and TJ Wilkason. PWG acknowledges the support of NSF grant PHY-1720397, DOE Early Career Award DE-SC0012012 and the W.M. Keck Foundation. The work of SJ is supported by the US Department of Energy contract DE-AC02-76SF00515, NRF Korea grant 2017R1D1A1B03030820, and Research Settlement Fund for the new faculty of Seoul National University. 

\end{acknowledgments}

\appendix
\section{GW Waveform} \label{app:waveform}

We collect the formula used in this paper here. The waveform in the frequency domain with spin-orbit coupling $\beta=0$~\cite{Cutler:1997ta,Cutler:1994ys,Nissanke:2009kt} is the Fourier-transform of the time-domain one in \Eq{eq:waveform-t}:  
\bea
\widetilde{h}(f) &=& \int_{-\infty}^{+\infty} h(t) \, e^{2\pi i f t} \, dt   \label{eq:waveform-f}  \\
 &=& \sqrt{\frac{5}{96}} \frac{\sqrt{A_+^2 F_+^2 + A_\times^2 F_\times^2}}{D_L} \, \pi^{-2/3} {\cal M}_z^{5/6} f^{-7/6} \, \exp[ i \Psi(f) ],
\eea
where the phase is
\beq
\Psi(f) \= 2\pi f t_c - \phi_c -\frac{\pi}{4} + \frac{3}{128} (\pi {\cal M}_z f)^{-5/3} - \phi_P(t) - \phi_D(t) \+ \cdots,
\eeq
and the chirp mass ${\cal M} = (M_1 M_2)^{3/5}/(M_1+M_2)^{1/5}$ with ${\cal M}_z \equiv {\cal M}(1+z)$. We do not write next-order Post-Newtonian corrections here, but we include them for phase in our numerical calculation~\cite{Cutler:1992tc}; see the above references for these next-order terms.
The Doppler phase measured by AI detectors contains important angular information as
\beq
\phi_D \= 2\pi f \, \big( \, R_{AI} {\bf r}_{AI} \cdot {\bf n}/c \+ R_{AU} {\bf r}_{Ea} \cdot {\bf n}/c \, \big)
\label{eq:dopplerphase} \eeq
with $R_{AU}=1$ AU, and the polarization phase is 
\beq 
\phi_P = \arctan[ (A_+ F_+)/(A_\times F_\times)]. 
\label{eq:polphase} \eeq 
Each polarization amplitude is $A_+ = 1+c_\iota^2$ and $A_\times = -2 c_\iota$.
The monotonic GW frequency evolution 
\beq
\frac{df}{dt} \= \frac{96}{5} \pi^{8/3} {\cal M}_z^{5/3} f^{11/3} \+ \cdots
\eeq
allows the one-to-one correspondence between GW frequency and the time before merger
\beq
t(f) \= t_c - \frac{5}{256} {\cal M}_z (\pi {\cal M}_z f)^{-8/3} \+ \cdots.
\eeq

We use the following source parameters for all benchmarks in this paper:
\beq
\theta \= \frac{\pi}{3.6}, \quad \phi \= \frac{\pi}{10.1}, \quad \psi \= \frac{\pi}{3}, \quad c_\iota \= \cos(150^\circ).
\label{eq:sourceparameters} \eeq
They are somewhat randomly chosen, but not leading to particularly good or bad angular resolution.

Polarization tensors used in \Eq{eq:hij_decomp} are
\beq
e^+_{ij} \= \hat{X}_i \hat{X}_j \- \hat{Y}_i \hat{Y}_j, \qquad e^\times_{ij} \= \hat{X}_i \hat{Y}_j \+ \hat{Y}_i \hat{X}_j,
\label{eq:polarizationtensor}\eeq
where the basis is~\cite{Cutler:1994ys,Nissanke:2009kt}
\bea
{\bf \hat{X}} &=& ( \, \sin \phi \cos \psi - \sin \psi \cos \phi \cos \theta, \, - \cos \phi \cos \psi - \sin \psi \sin \phi \cos \theta, \, \sin \psi \sin \theta \, ),  \\
{\bf \hat{Y}} &=& ( \, -\sin \phi \sin \psi - \cos \psi \cos \phi \cos \theta, \, \cos \phi \sin \psi - \cos \psi \sin \phi \cos \theta, \, \cos\psi \sin \theta \, ).
\eea

\section{Details of Fisher Calculation} \label{app:fisher}

Our Fisher matrix calculation involves a highly oscillatory integration due to the quick reorientation around the Earth. To facilitate Mathematica \texttt{NIntegrate} computation, we (1) divide integration frequency range (mostly according to the number of GW cycles), (2) keep relative errors of every subregion integral small and similar in size, and (3) use high-precision numerical variables. As a result, the inversion of the Fisher matrix becomes relatively stable (even though matrix condition numbers are very large).
But the calculation is more than 10 times slower than the calculation without the reorientation around the Earth. For an almost monochromatic GW (whose frequency evolves very slowly, e.g. in the inspiral phase far from merger), we use the following approximation to cross-check the full result of frequency-domain
\beq
\rho \simeq \frac{2}{S_n(f_0)} \int \, |h(t)|^2 \, dt,  \qquad \qquad \Gamma \simeq \frac{2}{S_n(f_0)} \int (\partial_i h ) (\partial_j h) \, dt.
\eeq

Using our AI calculation, we could approximately reproduce the LISA~\cite{Cutler:1997ta} and BBO~\cite{Crowder:2005nr,Cutler:2009qv} Fisher estimations of $\sqrt{\Omega_s}$ and $\Delta D_L/D_L$ normalized to a certain SNR from the last 1-yr observation. We take this as one cross-check of our calculation. This may also imply that, for long enough measurements, details of detector configuration and orbit are not so important in an order-of-magnitude estimation.

\section{Optimal separation of measurements} \label{app:discrete}

Given that the \emph{change} of Doppler shift contains measurable angular information, which two angles from a circular orbit can yield maximum angular information? By solving the $2 \times 2$ Fisher matrix $\Gamma_{2\times 2}$ composed of $\theta$ and $\phi$ (thus, ignoring any uncertainties correlated with other parameters), we obtain
\beq
\Delta \Omega_s \approx 2 \pi \sin \theta \, (\det \Gamma_{2\times 2})^{-1/2}.
\eeq
From the two measurements of $\delta$-duration($\delta \ll 1$ rad) separated by an orbit angle $\alpha$, the above $2 \times 2$ Fisher with Doppler effects only gives
\bea
\Delta \Omega_s^{-1} &\propto & (f R)^2 \sin 2\theta \sqrt{ 4 \delta^2 + \cos 2 \delta - 1 -2 \sin^2\delta \cos 2 \alpha } \nonumber\\
& \approx & (f R)^2 \sin 2\theta \sqrt{ 2 \delta^2 \left(1-\cos 2\alpha \right)}.
\eea
Thus, Doppler effects are maximized for  $\alpha \simeq \pi/2$. 
Locating two detectors separated by $\pi/2$ along the orbit, or measuring a GW at two different times separated by $\pi/2$ can thus maximize the Doppler effect. The former result is used to decide the locations of our two terrestrial detectors. The latter result can be useful in designing a measurement protocol of a narrow-band resonant-AI which has only limited measurement time for one GW.


\end{document}